\documentstyle[11pt]{article}
\def\fnote#1#2{\begingroup\def\thefootnote{#1}\footnote{#2}\addtocounter
{footnote}{-1}\endgroup}

\begin{document}

\hfill{UTTG-10-19}

\vspace{20pt}

\begin{center}
{\large {\bf { Models of Lepton and Quark Masses}}}

\vspace{20pt}

Steven Weinberg\fnote{*}{Electronic address:
weinberg@physics.utexas.edu}\\
{\em Theory Group, Department of Physics, University of
Texas\\
Austin, TX, 78712}

\vspace{30pt}

\noindent
{\bf Abstract}
\end{center}

A class of models is considered in which the masses only of the third generation of quarks and leptons arise in the tree approximation, while masses for the second and first generations are produced respectively by one-loop and two-loop radiative corrections.  So far, for various reasons, these models are not realistic.

\vfill

\pagebreak

\begin{center}
{\bf 1. Introduction}
\end{center}

In the Standard Model the masses of quarks and leptons take values proportional to the coupling constants in the interaction of these fermions with scalar fields, constants that in the context of this model are entirely arbitrary.  But the peculiar hierarchical pattern of lepton and quark masses seems to call for a larger theory, in which in some leading approximation the only quarks and leptons with non-zero mass are those of the third generation, the tau, top, and bottom, with the other lepton and quark masses arising from some sort of radiative correction.  Such theories were actively considered[1]	  soon after the completion of the Standard Model, but interest in this program seems to have lapsed subsequently[2].  

This paper will explore in detail a class of models of this sort, based on a different symmetry group.  These models are not realistic, for reasons that will be spelled out later,  but it is hoped that they may help to revive interest in this program, and to lay out some of the methods and problems that it confronts.

\begin{center}
{\bf 2. Gauge and Scalar Fields}
\end{center}

If the spontaneous breakdown of the electroweak symmetry gave masses only to the quarks and leptons of the third generation  in the tree approximation, then nothing in the Standard Model would generate masses for  the first and second generations in higher orders of perturbation theory.  To get masses for the second and first generations by emission and absorption of some sort of gauge bosons,  we would need to expand the gauge symmetry group.  In order for these masses to be much less than the zeroth order masses of the third generation, we would need the gauge coupling constants to be relatively small, more or less like the electroweak couplings.  If these new gauge couplings together  with those of the Standard Model all descended from some theory such as a string theory or a unified gauge theory in which they were all equal at some very high energy, then in order to have small couplings at accessible energies the new gauge group would have to be a direct product of simple subgroups with smaller beta functions than for the $SU(3)$ of QCD --- that is, most likely only $SO(3)$ and/or $SO(2)$.  After some attempts, what seems to work best is $SO_L(3)\otimes  SO_R(3)$, with the three generations of left-handed quark and lepton $SU(2)\otimes U(1)$  doublets forming  separate representations $(3,1)$ of $SO_L(3)\otimes  SO_R(3)$, and the three generations of right-handed quarks and charged leptons furnishing separate representations   $(1,3)$.  (We label representations of $SO(3)$ by their dimensionality.)  Though we shall concentrate on this gauge group, our analysis will deal with problems that would have to be encountered in any attempt to interpret the hierarchy of quark and lepton masses as radiative corrections.

In order for scalar fields to have renormalizable couplings to these quarks and leptons, they would have to form 9 electroweak doublets 
\begin{equation}
\left(\begin{array}{c} \Phi_{ia}^+\\ \Phi_{ia}^0\end{array}\right)\;,
\end{equation}
transforming as $(3,3)$ representations of $SO_L(3)\otimes  SO_R(3)$.  (Here superscripts indicate charges; subscripts $i$, $j$, etc are $SO_L(3)$ vector indices running over the values 1, 2, 3; subscripts $a$, $b$, etc are $SO_R(3)$ vector indices. also running over the values 1, 2, 3.)    Emission and absorption of the corresponding spinless particles also produces radiative corrections to the quark and lepton masses.  As we shall see in the next section, while keeping the mass of the Standard Model Higgs boson and the weak coupling constant at their known values, we can take all the other scalar particles and the new vector bosons to be  heavy enough to have escaped detection.  But the calculation in Section 4 show that the radiative corrections to masses do not disappear when the new scalar and vector bosons become very heavy.

The only possible renormalizable coupling of these scalars to leptons and quarks is then
\begin{eqnarray}
&&{\cal L}_{q\&\ell,\Phi}=-G_{\cal L}\overline{\left(\begin{array}{c}\nu_i\\ \ell^-_i\end{array}\right)_L}\cdot \left(\begin{array}{c}\Phi^+_{ia}\\\Phi^0_{ia}\end{array}\right)\ell^-_{Ra}\nonumber\\&&-G_D\overline{\left(\begin{array}{c}q^{2/3}_i\\ q^{-1/3}_i\end{array}\right)_L}\cdot \left(\begin{array}{c}\Phi^+_{ia}\\\Phi^0_{ia}\end{array}\right)q^{-1/3}_{Ra}
-G_U\overline{\left(\begin{array}{c}q^{2/3}_i\\ q^{-1/3}_i\end{array}\right)_L}\cdot \left(\begin{array}{c}\Phi^{0*}_{ia}\\\Phi^-_{ia}\end{array}\right)q^{+2/3}_{Ra}
\;+\;{\rm c.c.} \;.\nonumber\\&&{}
\end{eqnarray}
Here and below $G_{\cal L}$, $G_D$ and $G_U$ are  constants, and again $i$ and $a$ run over the values 1,2,3,  repeated indices are summed, and superscripts indicate charges.

\begin{center}
{\bf 3. Stationary Points: A First Look}
\end{center}

The most general renormalizable potential for the scalars $\Phi$ that is invariant under  the new $SO_L(3)\otimes  SO_R(3)$ as well as the electroweak $SU(2)\otimes U(1)$ takes the form
\begin{eqnarray}
&&V_I(\Phi)=-\mu^2\overline{\left(\begin{array}{c}\Phi^+_{ia}\\ \Phi^0_{ia}\end{array}\right)}\cdot \left(\begin{array}{c}\Phi^+_{ia}\\ \Phi^0_{ia}\end{array}\right)\nonumber\\&&
+
\overline{\left(\begin{array}{c}\Phi^+_{ia}\\ \Phi^0_{ia}\end{array}\right)}\cdot \left(\begin{array}{c}\Phi^+_{jb}\\ \Phi^0_{jb}\end{array}\right)\;
\overline{\left(\begin{array}{c}\Phi^+_{kc}\\ \Phi^0_{kc}\end{array}\right)}\cdot \left(\begin{array}{c}\Phi^+_{ld}\\ \Phi^0_{ld}\end{array}\right)\nonumber\\&& \times\Bigg[	
b_1\delta_{ij}\delta_{kl}\delta_{ab}\delta_{cd}+c_1\delta_{ij}\delta_{kl}\delta_{ac}\delta_{bd}+c_2\delta_{ij}\delta_{kl}\delta_{ad}\delta_{bc}\nonumber\\&&+c_3\delta_{ik}\delta_{jl}\delta_{ab}\delta_{cd}+b_2\delta_{ik}\delta_{jl}\delta_{ac}\delta_{bd}+c_4\delta_{ik}\delta_{jl}\delta_{ad}\delta_{bc}\nonumber\\&&+c_5\delta_{il}\delta_{jk}\delta_{ab}\delta_{cd}+c_6\delta_{il}\delta_{jk}\delta_{ac}\delta_{bd}+b_3\delta_{il}\delta_{jk}\delta_{ad}\delta_{bc}\Bigg]\;,
\end{eqnarray}
where the $b_n$ and $c_n$ are various real dimensionless constants.
The Lagrangian terms (2) and (3) along with the rest of the Lagrangian happen to be invariant under a reflection:
\begin{equation}
{\cal R}:~~~~~\Phi\rightarrow -\Phi~~~~~~q_L\rightarrow -q_L~~~~~~\ell_L\rightarrow -\ell_L\;,
\end{equation}
with right-handed fermions and all gauge fields left invariant.

We are concerned here only with stationary points of the potential for which charge is conserved, so in seeking such stationary points we set $\Phi_{ia}^+=0$.  Inspection of Eq.~(3) then shows that every term is symmetric between $\Phi^0$ and its Hermitian conjugate.  It follows that  if $V(\Phi^0)$ is stationary at a real value of $\Phi^0$ under variations that keep 
$\Phi^0$ real, then  at this point it is stationary under all variations of $\Phi^0$.  (In general, if $V(z)=V(z^*)$ then for $\lambda$ real   $V(\lambda+\epsilon)=V(\lambda+\epsilon^*)$ can have no terms of first order in ${\rm Im}\,\epsilon$.) 
We can therefore seek stationary points of the potential (not necessarily all stationary points) by taking the possible  expectation values $\phi_{ia}$ of $\Phi^0_{ia}$ to be real.  (Here and below, we use lower case letters to distinguish the possible spacetime-independent c-number expectation values of various scalar fields from the fields themselves.)
 For $\Phi^+_{ia}=0$ and $\phi_{ia}\equiv \Phi^0_{ia}$ real, the potential (3) must take the form of a  general renormalizable potential that is  invariant under $SO_L(3)\otimes  SO_R(3)$ and the reflection ${\cal R}$, and so 
\begin{equation}
V_I(\phi)=-\mu^2{\rm Tr}\Big(\phi^T\phi\Big)+b\Big[{\rm Tr}\Big(\phi^T\phi\Big)\Big]^2+c{\rm Tr}\Big(\phi^T\phi\phi^T\phi\Big)\;.
\end{equation}
The dimensionless constants $b$ and $c$ are linear combinations of the coefficients of the quartic terms in the general potential (3):  
\begin{equation}
 b=b_1+b_2+b_3\;,~~c=c_1+c_2+c_3+c_4+c_5+c_6 \;.
\end{equation}
(A trilinear  $SO_L(3)\otimes  SO_R(3)$-invariant term Det$\phi$ cannot arise from (3).  This can also be seen as a consequence of  invariance under the reflection (4).)

If this were the end of the story, and there were no other scalar fields with which the fields $\Phi_{ia}$ could interact, then Eq.~(5) would be the potential that governs the possible expectation values of these scalars.  We will have to introduce other scalar fields that do interact with the 
$\Phi_{ia}$, but it will be instructive first to consider the implications of the potential (5), returning later to consider the effect of interaction with other scalars.

To ensure that the  $SO_L(3)\otimes  SO_R(3)$ gauge symmetry is spontaneously broken at the stationary points of (5), we would need to take $\mu^2>0$.  In order for this potential to go to $+\infty$ rather than $-\infty$ when $\phi$ goes to infinity in any direction, the other constants in (5) would have to be in either range $
b>0\;\&\;c>-b$ or $c>0\;\&\; b>-c/3$,
or both.
Any $\phi$ can be diagonalized by an $SO_L(3)\otimes  SO_R(3)$ transformation, so we can characterize the various 
stationary points of the potential according to their  elements when diagonalized.  If we assume that $b>0$ and $-b<c<0$, then at the global minimum of the potential, the 
$\phi_{ia}$  when diagonalized would have two zero diagonal elements and one non-zero diagonal element $\lambda$,  with $ \lambda^2 =
 \mu^2/(2c+2b)\;,$ 
which we can define as the 33-element.  In this case naturally only the third generation of quarks and leptons would have masses in the tree approximation, given by
\begin{equation}
m_\tau=G_{\cal L}  \lambda\;,~~m_b=G_D  \lambda\;,~~m_t=G_U  \lambda\;.
\end{equation}
Just as in the Standard Model, the breaking of the electroweak symmetry gives masses to the $W$ and $Z$ and eliminates the Goldstone bosons associated with $\Phi^+_{33}$ 
and ${\rm Im}\,\Phi^0_{33}$, leaving a neutral scalar associated with 
${\rm Re}\,\Phi^0_{33}$ whose couplings to the third generation quarks and leptons are the same as for the Higgs boson of the Standard Model.  

This introduction of  new scalar doublets can be tolerated only if the masses of the new scalar particles introduced in this way can all be much larger than the Standard Model Higgs mass, $m_H=125$ GeV.  For the potential (5), the known value of $m_H$  fixes $\mu$  to have the value $m_H/2$, and the known coupling constant 
$G_F$ of the weak interaction fixes the expectation value $\lambda=\mu/\sqrt{2(b+c)}$ of the scalar field ${\rm Re}\Phi^0_{33}$ to have the value $2^{-1/4}G_F^{-1/2}=247$ GeV, so  $b+c$ would have to take  the value $b+c=\sqrt{2}G_Fm_H^2/8=0.032$, but $b$ and $c$ and all the other constants in Eq.~(3) would be otherwise unconstrained.  The squared masses of the spinless particles  associated with the real neutral scalar fields  Re$\Phi_{ia}^0$ (with $i\neq 3$ and $a\neq 3$) would all equal to  $-2\mu^2 c/(b+c)$, so these masses could be made reasonably large by taking $-c$ of order unity while keeping $b+c$ fixed.    (Recall that in order to make the stationary point with only ${\rm Re}\Phi^0_{33}$ non-zero the global minimum of the potential (5), we have assumed that $-b<c<0$.) In the absence of other scalar fields, the real neutral scalar fields  Re$\Phi_{i3}^0$ and Re$\Phi_{3a}^0$ (with $i\neq 3$ and $a\neq 3$) would be massless Goldstone bosons,  eliminated by the Higgs mechanism.    The masses of  the particles associated with the other scalars, ${\rm Im}\Phi^0_{ia}$ and $\Phi^+_{ia}$ with $i\neq 3$ or $a\neq 3$ would involve the many other constants in Eq.~(3), and could  presumably therefore be made arbitrarily large.    

Of course this is not the end of the story.  With nothing added to the model,  the  $SO_L(2)\otimes  SO_R(2)$ subgroup of $SO_L(3)\otimes  SO_R(3)$ with generators $t_{L3}$ and $t_{R3}$ would be unbroken; the two $SO_L(3)\otimes  SO_R(3)$ gauge bosons associated with this subgroup would be massless;  and symmetry under the reflections ${\cal R}\exp(i\pi t_{L2})$ and ${\cal R}\exp(i\pi t_{L1})$ would be unbroken, keeping the quarks and leptons of the first and second generations massless despite all radiative corrections.  We need to add a new sector of  scalar fields whose expectation values together with the primary sector expectation values $\phi_{ia}$ can break all of $SO_L(3)\otimes  SO_R(3)$ (or all but some finite subgroup), and allow for the  second and first generations of quarks and leptons to acquire  masses from one-loop and two-loop  radiative corrections.  With all scalar vacuum expectation values other than $\langle{\rm Re}\Phi^0_{33}\rangle$ taken very large, the $SO_L(3)\otimes  SO_R(3)$ gauge bosons would be all arbitrarily heavy.
(We will see that this does not eliminate contributions of radiative corrections to the quark and lepton masses.)  These new  scalar fields can be assumed to be hidden, in the sense that they are neutral under the electroweak gauge group, so that they have no renormalizable couplings to the quarks and leptons and do not introduce any mixing of the W  and Z with the $SO_L(3)\otimes  SO_R(3)$ gauge bosons.  But we will have to come back in Section 6 to see which other results of the present section survive the introduction of this hidden sector of scalar fields.

\begin{center}
{\bf 4.  Masses from Radiative Corrections}
\end{center}

Section 5 will offer some illustrative speculations regarding the nature of the scalar fields of the hidden sector, and the vector and scalar boson masses produced by their expectation values, but for the present we shall work with  general real symmetric mass-square  matrices for the $SO_L(3)\otimes  SO_R(3)$ vector bosons and for the various scalar bosons.   

The $6\times 6$ mass matrix of the $SO_L(3)\otimes  SO_R(3)$ vector bosons has six eigenvalues $\mu_n$, with six-component eigenvectors $\Big(u^{(n)}_{Li},u^{(n)}_{Ra}\Big)$, satisfying
\begin{eqnarray} 
&&\sum_j \mu^2_{Li,Lj}u^{(n)}_{Lj}+\sum_b \mu^2_{Li,Rb}u^{(n)}_{Rb}=\mu_n^2 u^{(n)}_{Li}\nonumber\\&&
\sum_j \mu^2_{Ra,Lj}u^{(n)}_{Lj}+\sum_b \mu^2_{Ra,Rb}u^{(n)}_{Rb}=\mu_n^2 u^{(n)}_{Ra}
\end{eqnarray}
These eigenvectors are orthogonal, and can be chosen real and orthonormal, so that
\begin{equation}
 \sum_i u^{(n)}_{Li}u^{(m)}_{Li}+\sum_a u^{(n)}_{Ra}u^{(m)}_{Ra}=\delta_{nm}
\end{equation}
These eigenvalues and eigenvectors are the ingredients we need in calculating the effects of emission and absorption of the vector bosons.

To one loop order, the emission and absorption of $SO_L(3)\otimes  SO_R(3)$ gauge bosons with an intermediate third-generation massive quark or lepton  gives the one-particle-irreducible  two-point function (in a classic notation) for quarks or leptons of the first or second generation with four-momentum $p^\nu$ in Feynman gauge:\footnote{Both indices on the two-point function $\Sigma$ in Eq.~(10) run over both left- and right-handed quark or lepton fields of the first and second generations, so in Eq.~(10) we do not bother to distinguish between  $SO_L(3)$  indices $i,j,k$ and $SO_R(3)$ indices that are elsewhere denoted $a, b, c$} 
\begin{eqnarray}
&& \Sigma_{ij}(p)=\frac{i}{4(2\pi)^4}\sum_{kln}\epsilon_{ki3}\epsilon_{lj3} \int \frac{d^4q}{[(p-q)^2+m_3^2-i\epsilon][q^2+\mu_n^2-i\epsilon]}\nonumber\\
&& \times \Bigg[g_L^2 u_{Lk}^{(n)}u_{Ll}^{(n)}\gamma^\mu (1+\gamma_5)[-i(p-q)^\nu\gamma_\nu+m_3]\gamma_\mu (1+\gamma_5)\nonumber\\
&&+g_R^2 u_{Rk}^{(n)}u_{Rl}^{(n)}\gamma^\mu(1-\gamma_5)[-i(p-q)^\nu\gamma_\nu+m_3]\gamma_\mu(1-\gamma_5)\nonumber\\
&& +g_L g_Ru_{Lk}^{(n)}u_{Rl}^{(n)}\gamma^\mu (1+\gamma_5)[-i(p-q)^\nu\gamma_\nu+m_3]\gamma_\mu (1-\gamma_5)\nonumber\\
&&+g_Lg_R u_{Rk}^{(n)}u_{Ll}^{(n)} \gamma^\mu (1-\gamma_5)[-i(p-q)^\nu\gamma_\nu+m_3]\gamma_\mu (1+\gamma_5)\Bigg]\;,
\end{eqnarray}
Here $g_L$ and $g_R$ are the gauge couplings of $SO_L(3)\otimes  SO_R(3)$, and 
the sums over $n$ run over the  six vector boson mass eigenvalues defined by Eq.~(8).  The generators of  $SO_L(3)$  and  $SO_R(3)$ are denoted $t_{Li}$ and $t_{Ra}$, and in the $(3,3)$ representation have the components $[t_{Lk}]_{ij}=i\epsilon_{kij}$ and $[t_{Rc}]_{ab}=i\epsilon_{cab}$.  Hence, for instance, the generators 
$t_{L1}$ and $t_{R1}$ produce transitions between the second and third generations.

We are interested in the case  in which in the tree-approximation only $m_3$ is non-zero, so to one loop order we can go on the mass shell for the first and second generations by simply setting $p^\nu=0$, in   which case (after discarding terms in the integrand odd in $q$)  the two-point function (10) takes the form
\begin{equation}
\Sigma_{ai}(0)=m_{ai}(1+\gamma_5)/2+m_{ia}(1-\gamma_5)/2\;,
\end{equation}
where 
\begin{equation}
m_{ai}=\frac{4ig_Lg_Rm_3}{(2\pi)^4}\sum_{kln}\epsilon_{ba3}\epsilon_{ji3}  u_{Rb}^{(n)}u_{Lj}^{(n)} \int \frac{d^4q}{[q^2+m_3^2-i\epsilon][q^2+\mu_n^2-i\epsilon]}\;.
\end{equation}

Each term in the sum over vector boson mass eigenvalues is logarithmically divergent, but the sum is convergent, because 
the completeness of the set of eigenvectors $u^{(n)}$ together with the orthonormality conditions (9) tell us that
$\sum_n u^{(n)}u^{(n)T}$ is the unit matrix, and in particular
\begin{equation}
\sum_n u_{Li}^{(n)}u_{Ra}^{(n)}=0
\end{equation}
The logarithmic divergences are independent of $\mu_n$, and so their sum is proportional to (13), and hence they cancel.  Indeed, as remarked by Barr and Zee[3], renormalizabilty makes this sort of  cancellation  inevitable, as there is no counterterm that could cancel an infinity.

After combining denominators, Wick rotating, integrating over $q^\mu$, and integrating over Feynman parameters, the mass matrix of the second and first generations is
\begin{equation}
m_{ai}=\frac{g_Lg_R m_3}{4\pi^2}\sum_{bjn}\epsilon_{ab3}\epsilon_{ij3}  u_{Rb}^{(n)}u_{Lj}^{(n)}\left[\frac{\mu_n^2\ln \mu_n^2-m_3^2\ln m_3^2}{\mu_n^2-m_3^2}
\right]\;.
\end{equation}
It makes no difference what units for mass we use in calculating the logarithms, since a change in units only gives a term proportional to the sum (13).
We can diagonalize the matrix $m_{ai}$ (which also gets rid of the $\gamma_5$s)   by multiplying  the left- and right-handed quark or lepton fields of the first and second generation with independent $2\times 2$  unitary matrices $U_L$ and $U_R$; the physical masses of the first and second generation quarks and leptons are then the  elements of the diagonal matrix $U_R^\dagger m U_L$.  

The couplings at zero momentum transfer of the field ${\rm Re}\Phi^0_{33}$ to the first and second generation of quarks and leptons would be generated by the same one-loop diagram, and would be the same as in the Standard Model.  At non-zero momentum transfer the coupling is modified by a form factor, but this form factor is   nearly  constant up to momentum transfers of order $m_3$ or the smallest $\mu_n$, whichever is greater.

This general class of models provides a plausible possible explanation of why the quarks and leptons of the first and second generations should be much less massive than their third generation counterparts, but so far we have seen no reason why the first generation should be so much lighter than the second.  But we can now easily describe the sort of vector  boson mass matrices that will give  masses for the second but not the first generation in one-loop order.  

If 
\begin{equation}
\mu^2_{L2,Ra}=0 ~~~~\&~~~~\mu^2_{Li,R2}=0
\end{equation}
for all $a$ and $i$, and if
\begin{equation}
\mu^2_{L2,Li}=0 ~~~~\&~~~~\mu^2_{R2,Ra}=0
\end{equation}
for all $i\neq 2$ and all $a\neq 2$, then obviously the only eigenvectors of $\mu^2$ with $L2$ or $R2$ components respectively have only $L2$ or $R2$ components, so 
 $u^{(n)}_{R2}=0$ for all eigenvectors $n$ for which 
$u^{(n)}_{Li}\neq 0$, and  $u^{(n)}_{L2}=0$  for all eigenvectors $n$ for which $u^{(n)}_{Ra}\neq 0$.  Inspection of Eq.~(14) shows then that to one-loop order, 
$m_{1i}=0$ and $m_{a1}=0$ for all $i$ and $a$.  The $2\times 2$ mass matrix $m_{ai}$ of the first and second generations in this order would then be already diagonal, with only one non-zero element, the  second generation mass $m_2=m_{22}$:
\begin{equation}
m_2=\frac{g_Lg_R m_3}{4\pi^2}\sum_{n}  u_{R1}^{(n)}u_{L1}^{(n)}\left[\frac{\mu_n^2\ln \mu_n^2-m_3^2\ln m_3^2}{\mu_n^2-m_3^2}
\right]\;.
\end{equation}
Repeating the same steps that led to Eq.~(17),  we see that the  first-generation quarks and leptons  get a two-loop mass
\begin{equation}
m_1=\frac{g_Lg_R m_2}{4\pi^2}\sum_{n}  u_{R3}^{(n)}u_{L3}^{(n)}\left[\frac{\mu_n^2\ln \mu_n^2-m_3^2\ln m_3^2}{\mu_n^2-m_3^2}\right]\;.
\end{equation}

It is easy to think of a  finite subgroup of $SO_L(3)\otimes SO_R(3 \otimes {\cal R}$ that if unbroken would ensure the validity of conditions (15) and (16) and thereby give vanishing first-generation masses in one-loop order.
(This unbroken subgroup  must be finite to avoid the appearance of new massless gauge bosons.)  We could take this unbroken symmetry as invariance under the operators
\begin{equation}
{\cal R}\exp\Big(i\pi t_{L1}\Big)~~~~~~\&~~~~~~ {\cal R}\exp\Big(i\pi t_{R1}\Big)
\end{equation}
where ${\cal R}$ is the reflection (4).  (This reflection has no effect on vector boson masses, but must be included in order for the appearance of the vacuum expectation value of ${\rm Re}\Phi_{33}^0$ not to break invariance under the transformations (19).)
Unfortunately, invariance under (19) would imply not only that conditions (15) and (16) are satisfied, so that the first generation quarks and leptons get no mass in one-loop order, but would also imply that 
$ \mu^2_{L3,Ra}=0$ and $\mu^2_{Li,R3}=0$ for all $i$ and $a$, which according to Eq.~(18) would imply also that the first generation quarks and leptons also get no mass in two-loop order.  Indeed, we could have seen this without looking into the details of radiative corrections.  Because ${\cal R}$ is defined to change the sign of all left-handed quark or lepton fields, the first  symmetry transformation (19) changes the sign of the left-handed first generation quark or lepton fields, so if this is an unbroken symmetry then the first generation does not get a mass from any source, including scalar boson as well as vector boson interactions.  

So where does the first generation get its masses?    
It is  possible that the first generation masses have nothing to do with vector boson emission and absorption.  It should be noted that the symmetry of the Lagrangian under the reflection  ${\cal R}$ is an accidental symmetry, in the sense that it is a consequence of the gauge symmetries of the theory and the condition of renormalizability.  It therefore need not be respected by operators in the Lagrangian of higher dimensionality, whose coefficients are suppressed by a negative power of some large mass, just as lepton conservation is not respected by dimension five operators added to the renormalizable Lagrangian of the Standard Model.  If ${\cal R}$ symmetry is violated in this way,  the first generation of quarks and leptons could get masses in the tree approximation from interactions of the fermion fields with two or more powers of 
scalar fields, masses that are small only because of the suppression of these non-renormalizable interactions.  But because the first generation quarks and leptons are much heavier than neutrinos while much lighter than the third generation quarks and leptons, the mass scale whose reciprocal appears in these higher dimensional operators would have to be much lighter than the mass scale in the interactions that give neutrinos their mass and much heavier than the third generation fermions.  

In the next section we take up a possibility that is more in the spirit of this paper, that for a suitable choice of a hidden sector of scalar fields, the potential accidently has a symmetry that unlike (19) is {\em not} a subgroup of the gauge group and ${\cal R}$, and which  has a subgroup that when unbroken naturally gives the vector boson mass matrix  the form required for radiative corrections to give masses to the second generation of quarks and leptons in one-loop order but to the first generation only in two-loop order.

Masses are also generated by radiative corrections due to emission and absorption of scalar bosons.  Here again we will keep to the general case in this section, leaving it for the next sections to consider specific forms for the mass matrix.

The charged scalar fields $\varphi_N^+$ that correspond to charged spinless particles of definite mass $M_{N+}$ are in general linear combinations of the previously introduced fields $\Phi_{ia}^+$:
\begin{equation}
\varphi^+_N=\sum_{ia} u^{(N+)}_{ia}\Phi^+_{ia}\;,
\end{equation}
with $u^{(N+)}_{ia}$ some constant coefficients found by diagonalizing the charged scalar mass matrix.  
Assuming again that only the third generation quarks get masses in the tree approximation, the one-loop two-point function  for quarks of the first and second generation is here of the same form as (11), except that here $m_{ai}$ is  complex:
\begin{equation}
\Sigma_{ai}(0)=m_{ai}(1+\gamma_5)/2+m^*_{ia}(1-\gamma_5)/2 \;.
\end{equation}
Again, we can diagonalize the matrix $m_{ai}$    by multiplying  the left- and right-handed quark or lepton fields of the first and second generation with independent $2\times 2$  unitary matrices $U_L$ and $U_R$, and  the physical masses of the first and second generation quarks and leptons are then the  elements of the diagonal matrix $U_R^\dagger m U_L$.

  Following the same methods that led to Eq.~(12), we find the one-loop contribution of charged scalar bosons to the masses of first- and second-generation quarks of charge $+2/3$
\begin{equation}
m^{+2/3}_{ai}=\frac{iG_UG_Dm_b}{(2\pi)^4}\sum_{N}  u_{3a}^{(N+)*}u_{i3}^{(N+)} \int \frac{d^4q}{[q^2+m_b^2-i\epsilon][q^2+M_{N+}^2-i\epsilon]}\;,
\end{equation}
and for first- and second-generation quarks of charge $-1/3$:
\begin{equation}
m^{-1/3}_{ai}=\frac{iG_UG_Dm_t}{(2\pi)^4}\sum_{N}  u_{3a}^{(N+)}u_{i3}^{(N+)*} \int \frac{d^4q}{[q^2+m_t^2-i\epsilon][q^2+M_{N+}^2-i\epsilon]}\;.
\end{equation}
Each term in these sums is logarithmically divergent, but the ultraviolet divergences again cancel in the sum.  To see this, it  is easiest to derive the necessary completeness relation by  requiring that the fields (20) of definite mass have a kinematic Lagrangian term: 
$$-\sum_N\partial_\mu\varphi^{+*}_N\partial^\mu\varphi^+_N\;,$$
so that the propagators of these fields have the conventional normalization that we assumed in deriving Eqs.~(22) and (23).  In order that this kinematic Lagrangian should agree with the correct kinematic Lagrangian $-\partial_\mu\Phi^{+*}_{ia}\partial^\mu\Phi^{+}_{ia}$, it is necessary that
$$
\sum_{N}  u_{ia}^{(N+)*}u_{jb}^{(N+)}=\delta_{ij}\delta_{ab}
$$
Eqs.~(22) and (23) were derived only for the first and second generation one-loop masses,  where both $i$ and $a$ equal 1 and/or 2, in which case this relation  gives
\begin{equation}
\sum_{N}  u_{3a}^{(N+)*}u_{i3}^{(N+)}=0\;
\end{equation}
The divergences in each term of Eqs.~(22) and (23) are independent of $N$, so the total divergence in the sums is proportional to (24), and hence vanishes.
Eqs.~(22) and (23) then give for the mass matrices of first- and second-generation quarks of charge $+2/3$ and $-1/3$:
\begin{equation}
m^{+2/3}_{ai}=\frac{G_UG_Dm_b}{16\pi^2}\sum_{N}  u_{3a}^{(N+)*}u_{i3}^{(N+)}\left[\frac{M_{N+}^2\ln M_{N+}^2-m_b^2\ln m_b^2}{M_{N+}^2-m_b^2}
\right] \;,
\end{equation}
\begin{equation}
m^{-1/3}_{ai}=\frac{G_UG_Dm_t}{16 \pi^2}\sum_{N}  u_{3a}^{(N+)}u_{i3}^{(N+)*}\left[\frac{M_{N+}^2\ln M_{N+}^2-m_t^2\ln m_t^2}{M_{N+}^2-m_t^2}\right]
 \;.
\end{equation}

The case of neutral scalars is more complicated, because the neutral fields  of definite mass are in general linear combinations of the neutral scalar fields of the hidden sector to be introduced in the next section, as well as of the fields $\Phi_{ia}^0$ introduced in Section 2 and their adjoints.  Separating the real and imaginary parts of any complex fields of definite mass, we can take all the neutral scalars of definite mass to be real, and write them as
\begin{equation}
  \varphi_N^0=\sum_{ia}\Big[u_{ia}^{(N0)} \Phi_{ia}^0+u_{ia}^{(N0)*} \Phi_{ia}^{0*}\Big]+\dots
\end{equation}
where the coefficients $u_{ia}^{(N0)}$ are various complex constants, and the dots indicate linear combinations of scalar fields of the hidden sector.
The mass matrix $m_{ia}$ appearing in the two-point function (21) for the first and second generation of leptons and quarks of each charge are then:
\begin{equation}
m^{\cal L}_{ai}=\frac{iG_{\cal L}^2m_\tau}{(2\pi)^4}\sum_{N}  u_{3a}^{(N0)}u_{i3}^{(N0)*} \int \frac{d^4q}{[q^2+m_\tau^2-i\epsilon][q^2+M_{N0}^2-i\epsilon]}\;,
\end{equation}
\begin{equation}
m^{-1/3}_{ai}=\frac{iG^2_Dm_b}{(2\pi)^4}\sum_{N}  u_{3a}^{(N0)}u_{i3}^{(N0)*} \int \frac{d^4q}{[q^2+m_b^2-i\epsilon][q^2+M_{N0}^2-i\epsilon]}\;,
\end{equation}
\begin{equation}
m^{+2/3}_{ai}=\frac{iG^2_Um_t}{(2\pi)^4}\sum_{N}  u_{3a}^{(N0)*}u_{i3}^{(N0)} \int \frac{d^4q}{[q^2+m_t^2-i\epsilon][q^2+M_{N0}^2-i\epsilon]}\;,
\end{equation}
Again, to deal with logarithmic divergences, we need a completeness relation.
We define these real neutral scalars so that the kinematic term
 in the Lagrangian is   
$$-\frac{1}{2}\sum_N\partial_\mu\varphi^{0}_N\partial^\mu\varphi^0_N\;,$$
In order that this should contain the correct kinematic term $-\partial_\mu\Phi_{ia}^{0*}\partial^\mu\Phi_{ia}^{0}$ for the neutral scalars introduced earlier, it is necessary that 
\begin{equation}
\sum_{N}  u_{ia}^{(N+)*}u_{jb}^{(N+)}=\delta_{ij}\delta_{ab}\,,~~~~
\sum_{N}  u_{ia}^{(N0)*}u_{jb}^{(N0)}=0
\;.
\end{equation}
In the case that concerns us here, in which both $i$ and $a$ are unequal to 3, the second relation tells us that
\begin{equation}
\sum_{N}  u_{3a}^{(N0)*}u_{i3}^{(N0)}=0
\end{equation}
so the logarithmic divergences cancel in Eqs.~(28)-(30), which give
\begin{equation}
m^{\cal L}_{ai}=\frac{G_{\cal L}^2m_\tau }{16 \pi^2}\sum_{N}  u_{3a}^{(N0)}u_{i3}^{(N0)*}\left[\frac{M_{N0}^2\ln M_{N0}^2-m_\tau^2\ln m_\tau^2}{M_{N0}^2-m_\tau^2}\right]
 \;,
\end{equation}
\begin{equation}
m^{-1/3}_{ai}=\frac{G^2_Dm_b}{16 \pi^2}\sum_{N}  u_{3a}^{(N0)}u_{i3}^{(N0)*}\left[\frac{M_{N0}^2\ln M_{N0}^2-m_b^2\ln m_b^2}{M_{N0}^2-m_b^2}\right]
 \;,
\end{equation}
\begin{equation}
m^{+2/3}_{ai}=\frac{G_U^2m_t}{16 \pi^2}\sum_{N}  u_{3a}^{(N0)*}u_{i3}^{(N0)}\left[\frac{M_{N0}^2\ln M_{N0}^2-m_t^2\ln m_t^2}{M_{N0}^2-m_t^2}\right]
 \;.
\end{equation}

Inspection of Eqs.~(25), (26), and (33)-(35) shows that in order for  one-loop scalar boson emission and absorption to give masses to the second generation of quarks and leptons but not the first generation, there would have to be a scalar field of definite mass that includes both $ \Phi_{32}$  and $ \Phi_{23}$ terms but none that contain  both $ \Phi_{31}$  and $ \Phi_{13}$ terms.  In this case for these radiative corrections to give masses to the first generation in two-loop order there would have to be scalar fields of definite mass that contain both $ \Phi_{12}$  and $ \Phi_{21}$ terms.  
We will have to wait until we return to the primary sector scalar fields in Section 6  to see whether these conditions are satisfied.

\begin{center}
{\bf 5. Hidden Sector Scalars}
\end{center}

To give masses only to the third generation of quarks and leptons in the tree approximation, we have assumed that the scalar fields ${\rm Re}\Phi^0_{ia}$ have non-vanishing expectation values only for $i=a=3$.  These break $SO_L(3)\otimes  SO_R(3)$ to the  $SO_L(2)\otimes  SO_R(2)$ subgroup with generators $t_{L3}$ and $t_{R3}$.  This symmetry breaking by itself gives non-vanishing values only for the following components of the $SO_L(3)\otimes  SO_R(3)$ vector boson mass-squared matrix: 
\begin{eqnarray*}
 &&\mu_{L1,L1}=\mu_{L2,L2}=g_L^2 \lambda^2\;,\\&&
\mu_{R1,R1}=\mu_{R2,R2}=g_R^2 \lambda^2\;.
\end{eqnarray*}
 where $\lambda=\langle{\rm Re}\Phi^0_{33}\rangle$.  

To produce additional components of the vector boson mass matrix, we   introduce  a number of  additional scalar field multiplets   that, like the electroweak doublet $\Phi_{ia}$, transform according to the $(3,3)$ representation of $SO_L(3)\otimes  SO_R(3)$, but unlike $\Phi_{ia}$ are neutral under the electroweak $SU(2)\otimes U(1)$, and therefore cannot have renormalizable interactions with the quarks and leptons.  
We will denote   these new electroweak-neutral multiplets as $\Psi^{(N)}_{ia}$ with $N\geq 1$.  
For simplicity, we assume that the Lagrangian is invariant under independent sign changes $\Psi^{(N)}\rightarrow -\Psi^{(N)}$ for each of the new scalar multiplets, as well as the reflection ${\cal R}$ and $SO_L(3)\otimes  SO_R(3)$.  Then the most general renormalizable potential for all  the scalars is
\begin{equation}
V=V_I+V_{II}+V'\;,
\end{equation}
where $V_I$ is given by Eq.~(3), $V_{II}$ is the most general renormalizable potential for the hidden sector scalars
\begin{eqnarray}
V_{II}=&&\sum_N\left[-\mu_N^2{\rm Tr}\Big(\Psi^{(N)T}\Psi^{(N)}\Big)+b_N\Big[{\rm Tr}\Big(\Psi^{(N)T}\Psi^{(N)}\Big)\Big]^2+c_N{\rm Tr}\Big(\Psi^{(N)T}\Psi^{(N)}\Psi^{(N)T}\Psi^{(N))}\Big)\right]\nonumber\\&&
+\frac{1}{2}\sum_{N\neq N'}\Bigg[\xi_{NN'} {\rm Tr}\Big(\Psi^{(N)T}\Psi^{(N)}\Psi^{(N')T}\Psi^{(N')}\Big)+\kappa_{N N'}{\rm Tr}\Big(\Psi^{(N)}\Psi^{(N)T}\Psi^{(N')}\Psi^{(N')T}\Big)\nonumber\\&&
+\zeta_{N N'} {\rm Tr}\Big(\Psi^{(N)T}\Psi^{(N')}\Psi^{(N)T}\Psi^{(N')}\Big)
+\rho_{N N'} {\rm Tr}\Big(\Psi^{(N)T}\Psi^{(N)}\Big){\rm Tr}\Big(\Psi^{(N')T}\Psi^{(N')}\Big)\nonumber\\&&+\sigma_{N N'} [{\rm Tr}\Big(\Psi^{(N)T}\Psi^{(N')}\Big)]^2\Bigg]\;,
\end{eqnarray}
and $V'$ is  the general interaction between the primary and hidden sectors:
\begin{eqnarray}
V'=&&\sum_N \xi_N \overline{\left(\begin{array}{c} \Phi^+_{ia}\\ \Phi^0_{ia}\end{array}\right)}\cdot\left(\begin{array}{c} \Phi^+_{ib}\\ \Phi^0_{ib}\end{array}\right)
\Psi^{(N)}_{jb}\Psi^{(N)}_{ja}\nonumber\\&&
+\sum_N \kappa_N \overline{\left(\begin{array}{c} \Phi^+_{ia}\\ \Phi^0_{ia}\end{array}\right)}\cdot\left(\begin{array}{c} \Phi^+_{ja}\\ \Phi^0_{ja}\end{array}\right)
\Psi^{(N)}_{ib}\Psi^{(N)}_{jb}\nonumber\\&&
+\sum_N \zeta_N \overline{\left(\begin{array}{c} \Phi^+_{ia}\\ \Phi^0_{ia}\end{array}\right)}\cdot\left(\begin{array}{c} \Phi^+_{jb}\\ \Phi^0_{jb}\end{array}\right)
\Psi^{(N)}_{ib}\Psi^{(N)}_{ja}\nonumber\\&&
+\sum_N \rho_N \overline{\left(\begin{array}{c} \Phi^+_{ia}\\ \Phi^0_{ia}\end{array}\right)}\cdot\left(\begin{array}{c} \Phi^+_{ia}\\ \Phi^0_{ia}\end{array}\right)
\Psi^{(N)}_{jb}\Psi^{(N)}_{jb}\nonumber\\&&
+\sum_N \sigma_N \overline{\left(\begin{array}{c} \Phi^+_{ia}\\ \Phi^0_{ia}\end{array}\right)}\cdot\left(\begin{array}{c} \Phi^+_{jb}\\ \Phi^0_{jb}\end{array}\right)
\Psi^{(N)}_{ia}\Psi^{(N)}_{jb}\;.
\end{eqnarray}
    Here 
the coefficients $\mu_N^2$, $b_N$, etc. are real but otherwise arbitrary, and summation of repeated subscripts is understood.

We again assume that the stationary point of the potential has $\Phi^+_{ia}=0$ and ${\rm Im}\Phi^0_{ia}=0$, so to find this stationary point we can lump ${\rm Re}\Phi^0_{ia}$ together with the $\Psi^{(N)}_{ia}$.  We write ${\rm Re}\Phi^0_{ia}$ as $\Psi_{ia}^{(0)}$, and express the total potential as a function of the  possible c-number expectation values $\psi^{(N)}$ with $N\geq 0$ of all these  scalar fields (again denoted with lower case letters):
\begin{eqnarray}
V=&&\sum_N\left[-\mu_N^2{\rm Tr}\Big(\psi^{(N)T}\psi^{(N)}\Big)+b_N\Big[{\rm Tr}\Big(\psi^{(N)T}\psi^{(N)}\Big)\Big]^2+c_N{\rm Tr}\Big(\psi^{(N)T}\psi^{(N)}\psi^{(N)T}\psi^{(N))}\Big)\right]\nonumber\\&&
+\sum_{N\neq N'}\Bigg[\xi_{NN'} {\rm Tr}\Big(\psi^{(N)T}\psi^{(N)}\psi^{(N')T}\psi^{(N')}\Big)+\kappa_{N N'}{\rm Tr}\Big(\psi^{(N)}\psi^{(N)T}\psi^{(N')}\psi^{(N')T}\Big)\nonumber\\&&
+\zeta_{N N'} {\rm Tr}\Big(\psi^{(N)T}\psi^{(N')}\psi^{(N)T}\psi^{(N')}\Big)
+\rho_{N N'} {\rm Tr}\Big(\psi^{(N)T}\psi^{(N)}\Big){\rm Tr}\Big(\psi^{(N')T}\psi^{(N')}\Big)\nonumber\\&&+\sigma_{N N'} [{\rm Tr}\Big(\psi^{(N)T}\psi^{(N')}\Big)]^2\Bigg]\;.
\end{eqnarray}
The sums now extend to $N=0$ and/or $N'=0$, with $\mu_0^2=\mu^2$, $b_0=b$, $c_0=c$, $\xi_{0N}=\xi_{N0}=\xi_N$, etc.

We note that each of the summed $SO_L(3)$ and $SO_R(3)$ vector indices occurs just twice in each term, so this potential has an 
accidental symmetry under the sign changes
\begin{equation}
\psi^{(N)}_{ia}\rightarrow \eta_{Li}\;\eta_{Ra}\;\psi^{(N)}_{ia}\;,
\end{equation}
where $\eta_{Li}$ and $\eta_{Ra}$ are any $N$-independent sign factors.  It is natural for the expectation values of the scalar fields to be invariant under any subgroup of this group of sign changes, in the technical sense that the restriction of the $\psi$s to such invariant values lowers the number of equations that need to be satisfied at a stationary point of the potential by the same amount as it lowers the number of free components of the $\psi$s.  In particular, it is natural to find stationary points that are invariant under the subgroup of the group of reflections (40) that consists of all the reflections with $
\eta_{L1}=-\eta_{L2}=-\eta_{L3}=\eta_{R1}=-\eta_{R2}=-\eta_{R3}$, or $\eta_{L2}=-\eta_{L1}=-\eta_{L3}=\eta_{R2}=-\eta_{R1}=-\eta_{R3}$, or $\eta_{L3}=-\eta_{L1}=-\eta_{L2}=\eta_{R3}=-\eta_{R1}=-\eta_{R2}$.  Invariance under this subgroup just implies that all $\psi^{(N)}_{ia}$ (including $\phi_{ia}\equiv \psi_{ia}^{(0)}$) are diagonal.   

At this  point we will greatly simplify  our discussion by taking the coefficient $\sigma_{NN'}$ of the final term in Eq.~(37) and the corresponding coefficient $\sigma_N$ in Eq.~(38) to vanish.  I have not been able to think of any symmetry assumption that would have this as a consequence, but setting all $\sigma_{NN'}$ and $\sigma_N$ equal to zero has the very convenient implication that with all $\psi^{(N)}_{ia}$  diagonal, the potential (39) is a function only of the {\em squares} of the diagonal components.  It is then natural to find a stationary point for which  in the basis in which all $\psi^{(N)}_{ia}$  are diagonal, there is  any desired assortment of zeroes on the diagonal of any $\psi^{(N)}_{ia}$.

Not only are such stationary points natural --- they all actually occur.  When the $\psi^{(N)}$ are all diagonal, the derivative of the potential (39) with respect to any one  component $\psi^{(N)}_{ia}$ takes the form
$$\frac{\partial V}{\partial \psi^{(N)}_{ia}}=\delta_{ia}\Big[-2\mu_N^2\psi^{(N)}_{ii}+\psi^{(N)}_{ii}\sum_{jN'}L_{Ni,N'j}[\psi^{(N')}_{jj}]^2\Big]\;,$$
where the $L$ are constants, depending on the coefficients $b_N$, $c_N$, $\xi_{NN'}$, etc. in the potential (39), but independent of the components of the $\psi$s.
(The summation convention is suspended here.)  The condition that $V$ be stationary with respect to variations in  $\psi^{(N)}_{ia}$ is trivially satisfied if $i\neq a$ or if $i=a$ and $\psi^{(N)}_{ii}=0$, while if $i=a$ and 
$\psi^{(N)}_{ii}\neq 0$ then this condition takes the form
$$2\mu_N^2=\sum_{jN'}L_{Ni,N'j}[\psi^{(N')}_{jj}]^2\;.$$
With a total of $D$ non-vanishing diagonal components, these are $D$ {\em linear} inhomogeneous equations for the squares of the $D$ non-vanishing components.  The determinant of $L$ does not vanish for generic values of the coefficients $b_N$, $c_N$, $\xi_{NN'}$, etc., so these equations have a solution, one that is unique. (We saw a simple example of this in Section 2.)  It is a more complicated business to find if this solution has positive values for all $[\psi^{(N')}_{jj}]^2$, and if this stationary point is an absolute minimum or even a local minimum of the potential.  In the absence of a specific candidate for a realistic theory, it does not seem worth while to go into this.

With diagonal  scalar expectation values, the $6\times 6$ vector boson mass-square matrix takes the block-diagonal form
\begin{equation}
\mu^2=\left(\begin{array}{ccc}\mu_1^2 & 0 & 0 \\0 & \mu_2^2 & 0 \\ 0 & 0 & \mu_3^2 \end{array}\right)
\end{equation}
where the $2\times 2$ submatrices $\mu_i^2$ are
\begin{equation}
\mu_i^2\equiv \left(\begin{array}{cc} \mu^2_{Li,Li} & \mu^2_{Li,Ri} \\ \mu^2_{Ri,Li} & \mu^2_{Ri,Ri}\end{array}\right)\;.
\end{equation}

To find a vector boson mass-square matrix of the sort that gives a hierarchy of quark and lepton masses, we can include  just three $(3,3)$ real neutral scalar multiplets, with non-zero expectation values:
\begin{eqnarray}
\psi^{(0)}=\left(\begin{array}{ccc}0 & 0 & 0 \\ 0 & 0 & 0 \\ 0 & 0 & \lambda\end{array}\right)\;,~~~
\psi^{(1)}=\left(\begin{array}{ccc}0 & 0 & 0 \\ 0 & \alpha & 0 \\ 0 & 0 & \beta\end{array}\right)\;,~~~
\psi^{(2)}=\left(\begin{array}{ccc}\gamma & 0 & 0 \\ 0 & \delta & 0 \\ 0 & 0 & 0\end{array}\right)\;.
\end{eqnarray}
  The vector boson mass-squared matrix here takes the block-diagonal form (41), (42), and now the non-vanishing elements of the submatrices are
\begin{eqnarray}
&&\mu_{L1,L1}^2=g_L^2(\lambda^2+\alpha^2+\beta^2+\gamma^2+\delta^2)\;,~~\mu_{L1,R1}^2=-2g_Lg_R\alpha\beta\;,\nonumber\\&&~~~~~~~~~~~\mu_{R1,R1}^2=g_R^2(\lambda^2+\alpha^2+\beta^2+\gamma^2+\delta^2)\;,\nonumber\\&&
\mu^2_{L2,L2}=g_L^2(\lambda^2+\beta^2+\gamma^2)\;,~~~\mu^2_{R2,R2}=g_R^2(\lambda^2+\beta^2+\gamma^2)\;,	\nonumber\\&&
\mu^2_{L3,L3}=g_L^2(\alpha^2+\gamma^2+\delta^2)\;,~~~\mu^2_{R3,R3}=g_R^2(\alpha^2+\gamma^2+\delta^2)\;,\nonumber\\&&
\mu^2_{L3,R3}=-2g_Lg_R\gamma\delta\;.
\end{eqnarray}
But note that $\mu^2_{L2,Ra}=\mu^2_{Li,R2}=0$ for all $i$  and $a$.
All the $SO_L(3)\otimes  SO_R(3)$ gauge bosons are massive, and can be made heavier than the $W$ and $Z$ by arranging that some combinations of $\alpha$, $\beta$, $\gamma$ and
$\delta$ (as for example just $\gamma$) are sufficiently larger than $\lambda$.  

 In one-loop order the emission of $L1$ and/or $R1$ gauge bosons in $2\leftrightarrow 3$ transitions produces second-generation masses given by Eq.~(33), but there still is no  mixing of the $L2$ and $R2$ gauge boson masses, so  one-loop radiative corrections  still do not give any  mass to the first  generation of quarks and leptons.  But because there now is a non-vanishing mixing of the $L3$ and $R3$ gauge boson masses, the emission of the  $L3$ component of a gauge boson in a $1\rightarrow 2$ transition followed by the absorption of the $R3$ component of the same gauge boson gives a mass for the first generation of quarks and leptons, which as shown in Eq.~(18) is proportional to the corresponding second-generation mass, and hence is of two-loop order.

\begin{center}
{\bf 6.  The Primary Sector Revisited}
\end{center}

It is easy to  preserve  the results of Section 2 when we add the hidden sector of scalar fields, by setting equal to zero all interactions of the scalar fields $\Phi_{ia}$ of the primary sector with the fields $\Psi^{(N)}_{ia}$ with $N\geq 1$ of the hidden sector --- that is, by setting the coefficients $\xi_N$, $\kappa_N$, $\zeta_N$, and $\rho_N$ in Eq.~(38) equal to zero (as well as taking all $\sigma_{N}$ to vanish).  But not only would this be an unnatural act; it would also have an unacceptable consequence.  With no interaction between the scalar fields of the primary and hidden sectors, the potential would be invariant under  {\em separate}   $SO_L(3)\otimes SO_R(3)$ transformations of the scalar fields of each sector.  When these two symmetries are spontaneously broken, there would be {\em two} sets of massless Goldstone bosons.  One linear combination of these massless fields would be eliminated by the Higgs mechanism, but since there is only one $SO_L(3)\otimes SO_R(3)$ gauge group, another linear combination would be left  as real massless spinless particles.  If we allow an interaction between the scalar fields of the primary and hidden sector, but assume that it is very weak, then the broken symmetry would entail a very light pseudo-Goldstone boson, which is almost as bad.  To avoid this, we must not only include the interactions in Eq ~(36) between the scalar fields of the primary and hidden sector, but also take these interactions strong enough to keep the pseudo-Goldstone boson too heavy to have been observed.  With this interaction present, it is necessary to reconsider the results in Section 3 for  the stationary points and masses of the scalar fields of the primary sector.

In carrying out this analysis, it is both necessary and convenient to assume that the expectation values of the scalar fields of the hidden sector are much larger than those of the primary sector.  This will ensure that the masses of the $SO_L(3)\otimes SO_R(3)$ gauge bosons are much larger than the W and Z masses.  The Goldstone boson that is eliminated by the Higgs mechanism is then close to the Goldstone boson associated with the symmetry breaking in the hidden sector, leaving a pseudo-Goldstone boson in the primary sector.  This assumption also gives the scalar fields of the hidden sector large masses, locking in their expectation values $\psi^{(N)}_{ia}$ with $N\geq 1$, independent of the fields of the primary sector.  

With this assumption, the potential for the scalar fields of the primary sector is effectively
\begin{equation}
V_{I, {\rm eff}}=V_I - \sum_{ia}\overline{\left(\begin{array}{c}\Phi_{ia}^+ \\ \Phi_{ia}^0\end{array}\right)}\cdot \left(\begin{array}{c}\Phi_{ia}^+ \\ \Phi_{ia}^0\end{array}\right)\mu_{ia}^2-\sum_{ia}\overline{\left(\begin{array}{c}\Phi_{ia}^+ \\ \Phi_{ia}^0\end{array}\right)}\cdot \left(\begin{array}{c}\Phi_{ai}^+ \\ \Phi_{ai}^0\end{array}\right)\mu'^2_{ia}\;,
\end{equation}
where 
\begin{equation}
\mu^2_{ia}=-\sum_{j,N\geq 1}\xi_N[\psi^{(N)}_{ja}]^2-\sum_{b,N\geq 1}\kappa_N[\psi^{(N)}_{ib}]^2-\sum_{b,j,N\geq 1}\rho_N[\psi^{(N)}_{jb}]^2
\end{equation}
\begin{equation}
\mu'^2_{ia}=\mu'^2_{ai}=-\sum_{N\geq 1}\zeta_N\psi^{(N)}_{ii}\psi^{(N)}_{aa};,
\end{equation}
and $V_I$ is given by Eq.~(3).

To find the stationary point of 
$V_{I, {\rm eff}}$, we again take the expectation values of $ \Phi^+_{ia} $ and  $ {\rm Im}\Phi^0_{ia}$ equal to zero, and let the expectation value $\phi_{ia}$ of    ${\rm Re}\Phi^0_{ia} $ have only the single non-zero element $\phi_{33}=\lambda$.  
(The existence of such a stationary point was shown in Section 5.)  Setting the term in Eq.~(45) of first order in $\Phi_{ia}^0-\langle \Phi_{ia}^0\rangle$ equal to zero then gives
\begin{equation}
\lambda^2=\frac{\mu^2+\mu_{33}^2+\mu'^2_{33}}{2(b+c)}\;,
\end{equation}
where $b$ and $c$ are given by Eq.~(6).  

The scalar mass matrix can be read off by considering the terms in Eq.~(45) that are quadratic in $\Phi_{ia}'\equiv \Phi_{ia}^0-\langle \Phi_{ia}^0\rangle$ and  $\Phi^+_{ia}$ and their adjoints:
\begin{eqnarray}
V_{I, {\rm quad}}&&=\sum_{i\neq 3, a\neq 3} [2\lambda^2 b_1-\mu^2-\mu^2_{ia}]\left|\Phi_{ia}^+\right|^2
+\sum_{i\neq 3}[2\lambda^2(b_1+c_1+c_2)-\mu^2-\mu_{i3}^2] \left|\Phi_{i3}^+\right|^2\nonumber\\&&+  \sum_{a\neq 3} [2\lambda^2(b_1+c_3+c_5)-\mu^2-\mu_{3a}^2]\left|\Phi_{3a}^+\right|^2-\sum_{i\neq 3, a\neq 3}\mu'^2_{ia}\Phi^{+\dagger}_{ia} \Phi^+_{ai} \nonumber\\&&
-2\sum_{i\neq 3}\mu'^2_{i3}\,{\rm Re}\Big(\Phi^{+\dagger}_{i3} \Phi^+_{3i}\Big)
\nonumber\\&& +[2\lambda^2(b+c)-\mu^2-\mu^2_{33}-\mu'^2_{33}]\left|\Phi_{33}^+\right|^2\nonumber\\
&&+\sum_{i\neq 3, a\neq 3} [2\lambda^2 b_1-\mu^2-\mu^2_{ia}]\left|\Phi_{ia}^0\right|^2\nonumber\\&&+\sum_{i\neq 3}[2\lambda^2(b_1+c_1+c_2+c_5+c_6+b_3)-\mu^2-\mu_{i3}^2] \left|\Phi_{i3}^0\right|^2\nonumber\\&&
+\sum_{a\neq 3}[2\lambda^2(b_1+c_3+c_5+c_2+c_4+b_3)-\mu^2-\mu_{3a}^2] \left|\Phi^0_{3a}\right|^2\nonumber\\&&
+[4\lambda^2(b+c)-\mu^2-\mu^2_{33}-\mu'^2_{33}]\left|\Phi'^0_{33}\right|^2+\sum_{i\neq 3, a\neq 3}2\lambda^2 b_3{\rm Re}\Big(\Phi_{ia}^0\Big)^2\nonumber\\&&
+\sum_{i\neq 3}2\lambda^2 (b_3+c_5+c_6){\rm Re}\Big(\Phi_{i3}^0\Big)^2+\sum_{a\neq 3}2\lambda^2 (b_3+c_2+c_4){\rm Re}\Big(\Phi_{3a}^0\Big)^2
\nonumber\\&&+2\lambda^2(b+c){\rm Re}\Big(\Phi'^0_{33}\Big)^2-\sum_{i\neq 3, a\neq 3}\mu'^2_{ia}\Phi_{ia}^{0\dagger}\,\Phi_{ai}^0\nonumber\\&&
-2\sum_{i\neq 3}\mu'^2_{i3}{\rm Re}\Big(\Phi_{i3}^{0\dagger}\Phi^0_{3i}\Big)
\;.
\end{eqnarray}

First, using Eq.~(48), we see that there are two fields here of zero mass: $\Phi_{33}^+$ and ${\rm Im}\Phi^0_{33}$.  These are the Goldstone bosons of broken $SU(2)\otimes U(1)$, and appear physically as the helicity zero states of the W and Z bosons, just as in the Standard Model.

Next, note that another  field of definite mass is ${\rm Re}\Phi_{33}'^0$, which plays the same role here as the Higgs boson of the Standard Model.
Its squared mass is the coefficient of $({\rm Re}\Phi_{33}'^0)^2/2$  in Eq.~(49).  Using Eq.~(48), this is
\begin{equation}
m_H^2=2[6\lambda^2(b+c)-\mu^2-\mu_{33}^2-\mu_{33}'^2]=4\Big(\mu^2+\mu_{33}^2+\mu_{33}'^2\Big)=8\lambda^2(b+c)
\end{equation}
The final result is the same as in Section 2, the only difference being that in the derivation $\mu^2$ is replaced with 
$\mu^2+\mu_{33}^2+\mu_{33}'^2$.  As noted in Section 2, our knowledge of the Higgs boson mass and the weak interaction strength lets us conclude from this result that $b+c$ has the value 0.032.  No other scalar boson mass is given by the same combination of parameters, so it is plausible that if the constants $b_n$ and $c_n$ in Eq.~(3) are of order unity then  all other scalar bosons are much heavier that the Higgs boson.

Of particular interest are the fields ${\rm Re}\Phi_{i3}^0$ and ${\rm Re}\Phi_{3i}^0$ with $i\neq 3$, which would be the Goldstone boson fields of broken 
$SO_L(3)\otimes SO_R(3)$ if the $\Phi^0_{ia}$ did not interact with the scalar fields of the hidden sector.  The terms in Eq.~(49) involving these fields are
\begin{eqnarray*}
&&\sum_{i\neq 3}\Bigg[\Big({\rm Re}\Phi^0_{i3}\Big)^2(2\lambda^2(b+c)-\mu^2-\mu_{i3}^2)
+\Big({\rm Re}\Phi^0_{3i}\Big)^2(2\lambda^2(b+c)-\mu^2-\mu_{3i}^2)\\&&
-2\Phi^0_{3i}\Phi^0_{i3}\mu'^2_{i3}\Bigg]\;,
\end{eqnarray*}
or, using Eq.~(48) again,
\begin{equation}
\sum_{i\neq 3}\Bigg[(\mu^2_{33}-\mu_{i3}^2)\Big({\rm Re}\Phi^0_{i3}\Big)^2
+(\mu^2_{33}-\mu_{3i}^2)\Big({\rm Re}\Phi^0_{3i}\Big)^2
-2\mu'^2_{i3}{\rm Re}\Phi^0_{3i}\,{\rm Re}\Phi^0_{i3}\Bigg]\;.
\end{equation}
As anticipated, these fields would evidently be massless  in the absence of the interaction between primary and hidden sectors.  
In the approximation assumed above in this section, that the breaking of $SO_L(3)\otimes SO_R(3)$ is mostly due to the expectation values of scalar fields of the hidden sector, the massless Goldstone boson fields associated with this breaking are dominated by terms $\Psi^{(N)}_{i3}$ and 
$\Psi^{(N)}_{3i}$ with $i\neq 3$ and $N\geq 1$, and it is these fields and not the primary sector fields ${\rm Re}\Phi^0_{i3}$ and ${\rm Re}\Phi^0_{3i}$that provide the helicity zero part of the massive  
$SO_L(3)\otimes SO_R(3)$ gauge bosons.

Continuing, the terms in  Eq.~(49) involving ${\rm Im}\Phi^0_{i3}$ and ${\rm Im}\Phi^0_{3i}$ with $i\neq 3$ are
\begin{eqnarray}
 &&\sum_{i\neq 3}\Bigg[[2\lambda^2(b_1+c_1+c_2)-\mu^2-\mu_{i3}^2]\Big({\rm Im}\Phi^0_{i3}\Big)^2+
 [2\lambda^2(b_1+c_3+c_5)-\mu^2-\mu_{3i}^2]\Big({\rm Im}\Phi^0_{3i}\Big)^2\nonumber\\&&~~~
-2\mu'^2_{i3}{\rm Im}\Phi^0_{i3}\,{\rm Im}\Phi^0_{3i}\Bigg]\;,
\end{eqnarray}
and the terms in Eq.~(49) involving ${\rm Re}\Phi^0_{ia}$ or ${\rm Im}\Phi^0_{ia}$ with $i\neq 3$ and $a\neq 3$ are
\begin{equation}
\sum_{i\neq 3,a\neq 3}\Bigg[[2\lambda^2(b_1+b_3)-\mu^2-\mu^2_{ia}]\Big({\rm Re}\Phi_{ia}^0\Big)^2-\mu'^2_{ia}{\rm Re}\Phi^0_{ia}\,{\rm Re}\Phi^0_{ai}\Bigg]
\end{equation}
and 
\begin{equation}
\sum_{i\neq 3,a\neq 3}\Bigg[[2\lambda^2(b_1-b_3)-\mu^2-\mu^2_{ia}]\Big({\rm Im}\Phi_{ia}^0\Big)^2-\mu'^2_{ia}{\rm Im}\Phi^0_{ia}\,{\rm Im}\Phi^0_{ai}\Bigg]\;.
\end{equation}
Corresponding results for the charged scalars can be found from the first four lines of Eq.~(49).

Inspection of these results shows that the charged and neutral scalar fields of definite mass contain both the terms $\Phi_{ia}$ and $\Phi_{ai}$ with $i\neq a$ if and only if $\mu'^2_{ia}$ is non-zero.  Eq.~(47) shows that for generic coefficients $\zeta_N$ this will be the case if there are one or more scalar fields $\Psi^{(N)}$ of the hidden sector whose expectation values $\psi^{(N)}$ have both $ii$ and $aa$ components non-zero.   
Eq.~(43) shows that for the choice we have made of the scalar fields of the hidden sector and for their expectation values, the coefficients $\mu'^2_{23}$ and $\mu'^2_{12}$ are non-zero, but $\mu'^2_{13}=0$.  It follows that there are  charged and neutral scalar fields of definite mass that contain both the terms $\Phi_{23}$ and $\Phi_{32}$, but none that contain both the terms $\Phi_{13}$ and $\Phi_{31}$.    As we have seen these are just the conditions  under which the quarks and leptons of the second generation but not the first generation get masses from one-loop emission and absorption of scalar bosons.  The one-loop  quark masses produced by charged scalars is given by Eq.~(25) and (26):
\begin{equation}
m_c=m^{+2/3}_{22}=\frac{G_UG_Dm_b}{16\pi^2}\sum_{N}  u_{32}^{(N+)*}u_{23}^{(N+)}\left[\frac{M_{N+}^2\ln M_{N+}^2-m_b^2\ln m_b^2}{M_{N+}^2-m_b^2}
\right] \;,
\end{equation}
\begin{equation}
m_s=m^{-1/3}_{22}=\frac{G_UG_Dm_t}{16 \pi^2}\sum_{N}  u_{32}^{(N+)}u_{23}^{(N+)*}\left[\frac{M_{N+}^2\ln M_{N+}^2-m_t^2\ln m_t^2}{M_{N+}^2-m_t^2}\right]
 \;,
\end{equation}
while the one-loop quark and lepton masses produced by neutral scalars are given by Eqs.~(33)-(35) as
\begin{equation}
m_\mu=m^{\cal L}_{22}=\frac{G_{\cal L}^2m_\tau }{16 \pi^2}\sum_{N}  u_{32}^{(N0)}u_{23}^{(N0)*}\left[\frac{M_{N0}^2\ln M_{N0}^2-m_\tau^2\ln m_\tau^2}{M_{N0}^2-m_\tau^2}\right]
 \;,
\end{equation}
\begin{equation}
m_s=m^{-1/3}_{22}=\frac{G^2_Dm_b}{16 \pi^2}\sum_{N}  u_{32}^{(N0)}u_{23}^{(N0)*}\left[\frac{M_{N0}^2\ln M_{N0}^2-m_b^2\ln m_b^2}{M_{N0}^2-m_b^2}\right]
 \;,
\end{equation}
and
\begin{equation}
m_c=m^{+2/3}_{22}=\frac{G_U^2m_t}{16 \pi^2}\sum_{N}  u_{32}^{(N0)*}u_{23}^{(N0)}\left[\frac{M_{N0}^2\ln M_{N0}^2-m_t^2\ln m_t^2}{M_{N0}^2-m_t^2}\right]
 \;.
\end{equation}
The first generation of quarks and leptons get masses from emission and absorption of scalar bosons in two-loop order.

It is striking that the same choice of scalar fields in the hidden sector leads to both the radiative corrections involving vector bosons and those involving scalar bosons generating  quark and lepton masses for the second and first generation  to one-loop and two-loop order, respectively.

\begin{center}
{\bf 7.   Problems}
\end{center}

The results obtained here for radiatively generated masses  involve many unknown parameters.  But we have noted that the large number of new scalar and vector particles in these models can (and must) be supposed to be heavy enough to have escaped observation, and where they are sufficiently heavy we can easily find the ratios of many  quark and lepton masses.   Unfortunately, these predicted ratios turn out to be wrong.

First,  if the masses of the second generation quarks and leptons arose from radiative corrections involving the $SO_L(3)\otimes  SO_R(3)$ gauge bosons, and if we did somehow arrange that these gauge bosons were all much heavier than the quarks and leptons of the third generations, then according to Eq.~(17) the ratios of the masses of the second and third generations  quarks and leptons would all be independent of the various masses of the third generation:
\begin{equation}
m_2/m_3\simeq \frac{g_Lg_R }{4\pi^2}\sum_{n}  u_{R1}^{(n)}u_{L1}^{(n)}\ln \mu_n^2
\;,
\end{equation}
 and so would be the same  for leptons and quarks of each charge, 
\begin{equation}
m_c/m_t=m_s/m_b=m_\mu/m_\tau
\end{equation}
which is not even approximately true of observed masses.  The strong interactions can account for some differences in these mass ratios, but these interactions are not very strong at energies of the order of $m_c$ and $m_t$, and of course are entirely absent for the leptons, while $m_c/m_t$ is an order of magnitude smaller than  $m_\mu/m_\tau$.  

Similarly, if the radiatively generated masses of second generation quarks were dominated by the emission and absorption of charged scalar bosons, then according to Eqs.~(55) and (56) we would have 
\begin{equation}
m_c/m_b=m_s/m_t
\end{equation}
which is even further from the truth.  Finally, if the radiatively generated masses of second generation quarks and leptons were dominated by the emission and absorption of neutral scalar bosons, then according to Eqs.~(57)--(59) and Eq.~(7) we would have 
\begin{equation}
m_c/m_t^3=m_s/m_b^3=m_\mu/m_\tau^3\;,
\end{equation}
which is worse yet.  

The above wrong predictions of mass ratios involving quarks of charge $+2/3$ would be invalidated if the mass of the relevant 
$SO_L(3)\otimes  SO_R(3)$ gauge bosons or scalar bosons were of the same order of magnitude of the top quark mass.  In that case, these new bosons might be accessible to observation.

Another unrealistic feature of these results is that they do not exhibit any Cabibbo-Kobayashi-Maskawa mixing angles.  Quark mass mixing could be included  by giving up the somewhat unnatural  assumption that the coefficients $\sigma_N$ and $\sigma_{NN'}$ in Eqs.~(38) and (39) all vanish.  
We might instead assume that for some reason these coefficients are relatively small, expecting that this will  yield rather small mixing angles.  But in this case it is not clear that it would be possible in a natural way to maintain the starting assumption that only the third generation of quarks and leptons get masses in the tree approximation.  Also, in this case we would need to worry about the possibility of flavor-changing neutral currents.

The best that can be hoped for the models discussed in this paper is that  they may perhaps provoke new ideas for  a realistic theory in which  radiative corrections  account for the masses of the first and second generations of quarks and leptons, together with  guidance in dealing with the problems that will arise in such a theory.

\begin{center}
 {\bf ACKNOWLEDGMENTS}
\end{center} 

I am grateful for conversations with Can Kilic and other members of the Theory Group at Austin, and for a correspondence with Anthony Zee.  This article is based on work supported by the National Science Foundation
 under Grants Number PHY-1620610 and PHY-1914679, and with support from the Robert A. Welch Foundation, Grant No. F-0014.

\begin{center}                               
---------------------------------------------------------
\end{center}                                                                                                                                                           \begin{enumerate}
\item S. M. Barr and A. Zee, Phys. Rev. D {\bf 17}, 1854 (1978); F Wilczek and A. Zee, Phys. Rev. Lett. {\bf 42}, 421 (1979); T. Yanagida, Phys. Rev. D {\bf 20}, 2986 (1979); and other references cited therein.
\item For an interesting exception, see B. A. Dobrescu and P. J. Fox, J. High Energy Phys. {\bf 8}, 1 (2008).
\item  S. M. Barr and A. Zee, ref. 1.  In this connection, also see S. Weinberg,   Phys. Rev. Lett. {\bf 29}, 388 (1972).
 \end{enumerate}

\end{document}